\documentclass[aps,prl,preprint,a4paper]{revtex4-1}

\usepackage{amsmath}
\usepackage{amssymb}
\usepackage{graphicx}
\usepackage{dcolumn}
\usepackage{bm}

\begin{document}


\title{Hamiltonian formulation of the gyrokinetic Vlasov-Maxwell equations} 



\author{J. W. Burby}
 \affiliation{Princeton Plasma Physics Laboratory, Princeton, New Jersey 08543, USA}
\author{A. J. Brizard}
 \affiliation{Department of Physics, Saint Michael's College, Colchester, Vermont 05439, USA}
\author{P. J. Morrison}
 \affiliation{Department of Physics and Institute for Fusion Studies, University of Texas at Austin, Austin, Texas 78712, USA}
\author{H. Qin}
 \affiliation{Princeton Plasma Physics Laboratory, Princeton, New Jersey 08543, USA}
 \affiliation{Dept. of Modern Physics, University of Science and Technology of China, Hefei, Anhui 230026, China}


\date{\today}

\begin{abstract}
The gyrokinetic Vlasov-Maxwell equations are cast as an infinite-dimensional Hamiltonian system. The gyrokinetic Poisson bracket is remarkably simple and similar to the Morrison-Marsden-Weinstein bracket for the Vlasov-Maxwell equations. Many of the bracket's Casimirs are identified. This work enables (i) the derivation of gyrokinetic equilibrium variational principles and (ii) the application of the energy-Casimir method and the method of dynamically-accessible variations to study stability properties of gyrokinetic equilibria.

\end{abstract}


\maketitle 

Most non-dissipative evolution equations in physics can be expressed in Hamiltonian form, often in terms of noncanonical variables. The well-known Hamiltonian formulation of Newton's equations of motion provides the most obvious finite-dimensional example\,\cite{Landau_1976}, but there are many interesting infinite-dimensional examples as well. For instance, Hamiltonian formulations of ideal magnetohydrodynamics and Vlasov-Maxwell dynamics were discovered by Morrison and Greene\,\cite{Morrison_MHD_1980} and Morrison, Marsden, and Weinstein \cite{Morrison_1980,Marsden_1982,Morrison_divB_1982}, respectively. Such Hamiltonian formulations provide access to specialized tools that lead to deep insights into a system's properties, such as equilibrium stability criteria\,\cite{Kruskal_Oberman_1958,Holm_stability_1985,Morrison_fluid_1998,Andreussi_2012,Andreussi_2013,Morrison_Pfirsch_1989} and the nature of the system's quantizations \cite{Dashen_1968,Wentzel_QFT_1949}. 

An important non-dissipative system that has \emph{never} been cast in Hamiltonian form is collisionless electromagnetic gyrokinetics. This theory, which was invented between 1968 and 1983 \cite{Taylor_1968,Rutherford_1968,Frieman_1982} in order to describe low-frequency microturbulence in plasmas, enjoys several \emph{Lagrangian} formulations. The earliest of these are given in Refs.\,\cite{Sugama_2000,Brizard_PRL_2000,Brizard_POP_2000}, while more recent additions can be found in Refs.\,\cite{Pfirsch_2004,Squire_GK_2013}. Given the typical intimate relationship between the Hamiltonian and Lagrangian 
formalisms, the Hamiltonian structure of electromagnetic gyrokinetics might therefore seem straightforward to obtain. This is not the case. The usual Legendre transform technique fails when applied to gyrokinetic Lagrangians. Nevertheless, the existence of Lagrangian formulations of electromagnetic gyrokinetics suggests that a Hamiltonian formulation \emph{should} exist; the riddle is how to find it.

In Ref.\,\cite{Squire_GK_2013} this Hamiltonian riddle was solved for \emph{electrostatic} gyrokinetics. The purpose of this Letter is to do the same for \emph{electromagnetic} gyrokinetics. Starting from a variational principle for the gyrokinetic Vlasov-Maxwell equations, we have systematically derived a gyrokinetic Vlasov-Maxwell Poisson bracket and Hamiltonian functional. We will present the results of this derivation and list many of the Casimir invariants of the gyrokinetic Poisson bracket. Details of the derivation will be found in a forthcoming article.

The gyrokinetic Vlasov-Maxwell equations are most easily defined by specifying their Lagrangian, $L_{\text{GMV}}$, which has the general form
\begin{align}
L_{\text{GMV}}=L_{\text{M}}+L_{\text{gy}},
\end{align}
where  $L_{\text{M}}$ is the free electromagnetic field Lagrangian and $L_{\text{gy}}$ is the net \emph{gyrocenter}\,\cite{Brizard_2007} Lagrangian.

The free electromagnetic field Lagrangian is given by the well-known expression \cite{Landau_fields_1975}
\begin{align}
L_{\text{M}}=\frac{1}{8\pi}\int_QE\wedge*E-B\wedge*B.
\end{align}
Here $Q$ is the $3$-dimensional Euclidean space, $E$ is the fluctuating electric field $1$-form, $B$ is the fluctuating magnetic field $2$-form, and $*$ is the hodge star on $Q$. $E$ and $B$ are related to the electric field and magnetic field \emph{vectors}, $\bm{E}$ and $\bm{B}$, by the relations
\begin{align}
E&=\bm{E}\cdot dx\\
B&=\bm{B}\cdot dS.
\end{align}
For a treatment of the Morrison-Marsden-Weinstein Vlasov-Maxwell bracket in terms of differential forms, see Ref.\,\cite{Vittot_2014}. 

The net gyrocenter Lagrangian has many possible expressions, owing to the fact that there are many different \emph{representations}\,\cite{Brizard_2012,Burby_gc_2013} of gyrocenter dynamics. By using one of the gauge-invariant representations introduced in Ref.\,\cite{Pfirsch_2004}, the net gyrocenter Lagrangian takes the form
\begin{align}
L_{\text{gy}}&=L_{\text{f}}+L_{\text{int}},
\end{align}  
where $L_{\text{f}}$ is the \emph{free gyrocenter Lagrangian} and $L_{\text{int}}$ is the \emph{interaction Lagrangian}.

The free gyrocenter Lagrangian is given by
\begin{align}
L_{\text{f}}=\sum_{s=1}^{N_s}\int_{TQ}(\Xi_s^{\text{gc}}(V_s)-K_s(E,B))\,f_s.
\end{align} 
Here $N_s$ is the number of plasma species, $TQ$ is the tangent bundle over $Q$, $\Xi_s^{\text{gc}}$ is the \emph{guiding} center Lagrange $1$-form\,\cite{Cary_2009} (also see Eq.\,(33) in Ref.\,\cite{Burby_gc_2013}), $V_s$ is the Eulerian phase space velocity of the $s$'th species' phase space fluid, $f_s$ is the gyrocenter phase space density $6$-form, and $K_s(E,B)$ is what we will refer to as the gyrocenter Kinetic energy. We work in terms of the gyrocenter phase space density $6$-form instead of the usual gyrocenter distribution \emph{function} because it simplifies the ensuing discussion. \emph{After a phase space volume form}, $\Omega$, \emph{is chosen}, the gyrocenter distribution function, $F$, can be recovered from the $6$-form using the formula $f=F\Omega$ \footnote{Note that the definition of the distribution function depends on one's (arbitrary) choice of volume form on phase space, whereas the $6$-form is not; regardless of which $\Omega$ is chosen, the number of particles of species $s$ in a region $R\subset TQ$ is simply $\int_{R}f_s$.}. The introduction of the gyrocenter kinetic energy $K_s(E,B)$, which is a non-local functional of the electromagnetic fields, is motivated by the work of Morrison in Ref.\,\cite{Morrison_lifting_2013}. The only property of $K_s$ we will need in order to describe the Hamiltonian formulation of the gyrokinetic Vlasov-Maxwell equations is that it functionally depends on the fields $E,B$ and \emph{not} directly on the potentials $\phi,A$. An expression for $K_s(E,B)$ will be given after presenting the gyrokinetic Vlasov-Maxwell Poisson bracket.

The interaction Lagrangian is given by
\begin{align}\label{l_int}
L_{\text{int}}=\sum_{s=1}^{N_s}\int_{TQ}\left(\frac{e_s}{c}(\pi^*A)(V_s)-e_s\pi^*\phi\right)\,f_s,
\end{align}
where $A$ is the vector potential $1$-form, $\phi$ is the scalar potential, and $\pi:TQ\rightarrow Q$ is the tangent bundle projection, $\pi(x,v)=x$. The electric field $1$-form and magnetic field $2$-form are given in terms of the potentials by $B=\mathbf{d}A$ and $E=-\mathbf{d}\phi-\dot{A}/c$. The presence of the pullback operator $\pi^*$ is necessitated by the fact that the integrand in Eq.\,(\ref{l_int}) must live on $TQ$, whereas $A$ and $\phi$ live on $Q$. If the standard gyrokinetic coordinates, $(X,\mu,v_\parallel,\theta)$, are used on $TQ$, $(\pi^*A)(X,\mu,v_\parallel,\theta)=A_i(X)\mathbf{d}X^i$ and $(\pi^*\phi)(X,\mu,v_\parallel,\theta)=\phi(X)$. Note that the only explicit appearance of the potentials $A,\phi$ in $L_{\text{gk}}$ is made in $L_{\text{int}}$.

The gyrokinetic Vlasov-Maxwell equations follow from the Lagrangian $L_{\text{GMV}}=L_{\text{f}}+L_{\text{int}}+L_{\text{M}}$ and the condition
\begin{align}
\delta\int_{t_1}^{t_2}L_{\text{GMV}}\,dt=0,
\end{align}
where $\delta$ indicates varying the quantities $A,\phi,f_s,V_s$ while keeping their values at $t_1$ and $t_2$ fixed. The variations of $f_s$ and $V_s$ are \emph{constrained}\,\cite{Squire_GK_2013,Holm_1998} to be of the form
\begin{align}
\delta V_s&=\dot{\eta}_s+L_{V_s}\eta_s\\
\delta f_s&=-L_{\eta_s}f_s,
\end{align}
where $\eta_s$ is an arbitrary vector field on $TQ$ and $L_{V_s}$ denotes the Lie derivative along $V_s$. Thus, the gyrokinetic Vlasov-Maxwell equations \footnote{The gyrokinetic Vlasov-Maxwell equations differ from the most common versions of the electromagnetic gyrokinetic equations. However, this difference is mainly due to the fact that we have retained small terms in the free electromagnetic Lagrangian that are typically dropped. } are given by
\begin{subequations}
\label{gvm}
\begin{align}
\frac{\partial}{\partial t}f_s&=-L_{V_s^{\text{gy}}}f_s\label{gk_vlasov}\\
\frac{1}{c}\frac{\partial}{\partial t}D&=\bm{\delta}H-\frac{4\pi}{c}J_{\text{gy}}\label{gk_ampere}\\
\frac{1}{c}\frac{\partial}{\partial t}B&=-\mathbf{d}E\label{gk_faraday}\\
\bm{\delta}D&=-4\pi\rho_{\text{gy}}\label{gk_gauss}\\
\mathbf{d}B&=0,\label{gk_divb}
\end{align}
\end{subequations}
where $\bm{\delta}$ is the codifferential\,\cite{FoM} (also see\,\footnote{The codifferential is the adjoint of the exterior derivative relative to the inner product of forms given by $\left<\alpha,\beta\right>=\int_Q\alpha\wedge*\beta$. Thus, if $\beta$ is a $k+1$-form, $\bm{\delta}\beta$ is the unique $k$-form that satisfies $\left<\alpha,\bm{\delta}\beta\right>=\left<\mathbf{d}\alpha,\beta\right>$ for each $k$-form $\alpha$.}), and we have introduced the gyrocenter Eulerian phase space velocity $V_s^{\text{gy}}$, the gyrocenter current density $1$-form $J_{\text{gy}}$, the gyrocenter charge density $\rho_{\text{gy}}$, and the auxiliary fields $D$ and $H$. The gyrocenter Eulerian phase space velocity is defined in terms of the gyrocenter symplectic form, $\omega_s^{\text{gy}}=-\mathbf{d}(\Xi_s^{\text{gc}}+e_s\pi^*A/c)$, according to
\begin{align}\label{velocity}
\text{i}_{V_s^{\text{gy}}}\omega_s^{\text{gy}}=\mathbf{d}K_s+e\pi^*E.
\end{align}
Essentially, $V_s^{\text{gy}}$ is the Lorentz force vector field, $v\cdot\partial_x+(e/m)(\bm{E}+v\times(\bm{B}+\bm{B}_o)/c)\cdot\partial_v$, expressed in gyrocenter coordinates. See Eq.\,(3.19) in Ref.\,\cite{Cary_2009} for the leading-order contribution to $\omega_s^{\text{gy}}$, which must be inverted to solve for $V_s^{\text{gy}}$ using Eq.\,(\ref{velocity}). The auxiliary fields are given by the constitutive relations
\begin{align}
D&=E-4\pi\frac{\delta\mathcal{K}}{\delta E}\\
H&=B+4\pi\frac{\delta\mathcal{K}}{\delta B},
\end{align}
where $\mathcal{K}(f,E,B)=\sum_{s=1}^{N_s}\int_{TQ}K_s(E,B)\,f_s$.
Note that these expressions imply that the gyrocenter polarization and magnetization are given by $P=-\delta\mathcal{K}/\delta E$ and $M=\delta\mathcal{K}/\delta B$, respectively, just as in Ref.\,\cite{Morrison_lifting_2013}. The gyrocenter current density $1$-form, $J_{\text{gy}}$, is related to the gyrocenter current density vector, $\bm{J}_{\text{gy}}$, by the relation
\begin{align}
J_{\text{gy}}=\bm{J}_{\text{gy}}\cdot dx.
\end{align}
Note that because $\bm{\delta}D=-\text{div}\bm{D}$ and $\bm{\delta}H=\text{curl}(\bm{H})\cdot dx$, where $\bm{D}$ and $\bm{H}$ are the vector auxiliary fields, these equations are identical to Maxwell's equations in a polarized and magnetized medium coupled with a kinetic equation for the ``free charge", which is naturally the gyrocenter charge.

Equations\,(\ref{gk_vlasov})-(\ref{gk_divb}) constitute an infnite-dimensional Hamiltonian system. In order to demonstrate this, we follow the example set by Morrison \cite{Morrison_lifting_2013} and use $f_s$, $D$, and $B$ as dynamical variables. The relevant Poisson bracket and Hamiltonian functional on $(f,D,B)$-space (where $B$ is required to satisfy $\mathbf{d}B=0$) are given as follows. 
\\ \\
\emph{The Poisson bracket}\,---
The gyrokinetic Vlasov-Maxwell Poisson bracket is given by
\begin{align}\label{the_bracket}
&[\mathcal{F},\mathcal{G}]_{\text{GVM}}=\nonumber\\
&\sum_{s=1}^{N_s}\int_{TQ}\mathcal{B}_{s}^{\text{gy}}\left(\mathbf{d}\frac{\delta\mathcal{F}}{\delta f_s}-4\pi e_s\pi^*\frac{\delta\mathcal{F}}{\delta D},\mathbf{d}\frac{\delta\mathcal{G}}{\delta f_s}-4\pi e_s\pi^*\frac{\delta\mathcal{G}}{\delta D}\right)\,f_s\nonumber\\
&+4\pi c\int_Q\frac{\delta\mathcal{F}}{\delta D}\wedge*\bm{\delta}\frac{\delta\mathcal{G}}{\delta B}-\frac{\delta\mathcal{G}}{\delta D}\wedge*\bm{\delta}\frac{\delta\mathcal{F}}{\delta B}.
\end{align}   
Here $\mathcal{B}_s^{\text{gy}}$ is the gyrocenter Poisson tensor, which is defined as follows. If $\alpha$ is a $1$-from on phase space and $(\omega_{s}^{\text{gy}})^{-1}(\alpha)=X$ is the unique vector field on phase space that satisfies
\begin{align}
\text{i}_{X}\omega_s^{\text{gy}}=\alpha,
\end{align}
then the gyrocenter Poisson tensor is given by
\begin{align}
\mathcal{B}_s^{\text{gy}}(\alpha_1,\alpha_2)=\omega_s^{\text{gy}}\left((\omega_{s}^{\text{gy}})^{-1}(\alpha_1),(\omega_{s}^{\text{gy}})^{-1}(\alpha_2)\right).
\end{align}
Note that if $\alpha_i=\mathbf{d}h_i$ for functions $h_i$, $\mathcal{B}_s^{\text{gy}}(\alpha_1,\alpha_2)=\{h_1,h_2\}_s^{\text{gy}}$, where $\{\cdot,\cdot\}_s^{\text{gy}}$ is the gyrocenter Poisson bracket. See Eq.\,(3.29) in Ref.\,\cite{Cary_2009} for the leading-order contribution to the gyrocenter Poisson bracket. The method used to derive the bracket given in Eq.\,(\ref{the_bracket}), which we will report on in a future publication, guarantees that it satisfies the Jacobi identity. 
\\ \\
\emph{The Hamiltonian functional}\,--- The gyrokinetic Vlasov-Maxwell Hamiltonian functional is given by
\begin{align}\label{gvm_hamiltonian}
\mathcal{H}_{\text{GVM}}(f,D,B)&=\mathcal{K}(f,\hat{E},B)+\int_Q\hat{P}\wedge*\hat{E}\nonumber\\
&+\frac{1}{8\pi}\int_Q\hat{E}\wedge*\hat{E}+B\wedge *B,
\end{align}
where $\hat{E}=\hat{E}(f,D,B)$ is the electric field operator defined implicitly by the equation
\begin{align}
D&=\hat{E}(f,D,B)-4\pi\frac{\delta\mathcal{K}}{\delta E}(f,\hat{E}(f,D,B),B),
\end{align}
and $\hat{P}=\hat{P}(f,D,B)$ is the gyrocenter polarization operator given by
\begin{align}
\hat{P}(f,D,B)=\frac{1}{4\pi}(D-\hat{E}(f,D,B)).
\end{align}
The functional derivatives of $\mathcal{H}_{\text{GVM}}$ are given by
\begin{subequations}
\begin{align}
\frac{\delta\mathcal{H}_{\text{GVM}}}{\delta f_s}(f,D,B)&=K_s(\hat{E},B)\\
\frac{\delta\mathcal{H}_{\text{GVM}}}{\delta D}(f,D,B)&=\frac{1}{4\pi}\hat{E}\\
\frac{\delta\mathcal{H}_{\text{GVM}}}{\delta B}(f,D,B)&=\frac{1}{4\pi}\left(B+4\pi\frac{\delta\mathcal{K}}{\delta B}(f,\hat{E},B)\right).
\end{align}
\end{subequations}
It is straightforward to verify that the dynamics on $(f,D,B)$-space defined by 
\begin{align}
\dot{\mathcal{Q}}=[\mathcal{Q},\mathcal{H}_{\text{GVM}}]_{\text{GVM}},
\end{align}
where $\mathcal{Q}$ is an arbitrary functional of $(f,D,B)$, reproduce the gyrokinetic Vlasov-Maxwell equations (note that, by the gyrocenter continuity equation, Eq.\,(\ref{gk_gauss}) is automatically satisfied for all times if it is satisfied at $t=0$.)

A large class of Casimirs of the gyrokinetic Vlasov-Maxwell bracket are given as follows. Let $\mathbb{C}(\kappa,\beta,\lambda)$ be a real-valued functional depending on (i) a function on $Q$, $\kappa$, (ii) a $2$-form on $TQ$, $\beta$, and (iii) a $6$-form on $TQ$, $\lambda$. Suppose that $\mathbb{C}$ satisfies the invariance property
\begin{align}
\mathbb{C}(\kappa,\Phi^*\beta,\Phi^*\lambda)=\mathbb{C}(\kappa,\beta,\lambda),
\end{align}
for each diffeomorphism of $TQ$, $\Phi:TQ\rightarrow TQ$. Then the functional
\begin{align}
C(f,D,B)=\mathbb{C}(\bm{\delta}D+4\pi\rho_{\text{gy}},\omega^{\text{gy}},f)
\end{align}
is a Casimir. In particular, if we let
\begin{align}
\Omega_s=-\frac{1}{3!}\omega_s^{\text{gy}}\wedge\omega_s^{\text{gy}}\wedge\omega_s^{\text{gy}}
\end{align}
be the Liouville volume form defined by the gyrocenter symplectic form and introduce the gyrocenter distribution \emph{function}, $F_s$, where
\begin{align}
f_s=F_s\Omega_s,
\end{align} 
then
\begin{align}
C_h=\sum_{s=1}^{N_s}\int_{TQ}h(F_s)\,\Omega_s
\end{align}
is a Casimir for each function of a single real variable $h$. Moreover, any functional of $\bm{\delta}D+4\pi\rho_{\text{gy}}$ is a Casimir, which is one way of seeing that Eq.\,(\ref{gk_gauss}) is satisfied in the Hamiltonian formulation of the gyrokinetic Vlasov-Maxwell equations.

Using the knowledge of the form of the Poisson bracket $[\cdot,\cdot]_{\text{GVM}}$ along with its Casimirs, equilibrium variational principles and stability criteria can be formulated as in Refs.\,\cite{Kruskal_Oberman_1958,Holm_stability_1985,Morrison_fluid_1998,Andreussi_2012,Andreussi_2013}. The equilibrium variational principles are based on the fact that, for each Casimir $C$, critical points of the free energy functional
\begin{align}
F=\mathcal{H}_{\text{GVM}}+C
\end{align}
are gyrokinetic equilibria. Any equilibrium that can be obtained in this manner will be stable (linearly, and often times non-linearly) if the second variation of $F$ is positive definite. Additionally,  knowledge of the Poisson bracket allows one to construct the dynamically accessible variations introduced in Ref.\,\cite{Morrison_Pfirsch_1989} (see also \cite{Morrison_fluid_1998,Andreussi_2013}).  These constrained variations yield all equilibria as energy extrema, and upon second order constrained variation the linearized energy, a quadratic form, definiteness of which implies stability.

The Hamiltonian formulation of the gyrokinetic Vlasov-Maxwell equations given in this Letter is completely determined by two key quantities, the gyrocenter kinetic energy, and the guiding center Lagrange $1$-form. Suppressing species labels, the gyrocenter kinetic energy is given explicitly to second order in the amplitude of the fluctuating fields, $\epsilon_\delta$, by
\begin{align}
&K(E,B)=\mathcal{H}^{\text{gc}}-\epsilon_\delta \left<\ell\right>+\epsilon_\delta ^2\mathcal{B}^{\text{gy}}(\left<\delta\Xi\right>,\mathbf{d}\left<\ell\right>)\nonumber\\
                &+\frac{1}{2}\epsilon_\delta^2\left<\mathcal{B}_s^{\text{gy}}\bigg(L_R[\delta\tilde{\Xi}-\mathbf{d}I(\tilde{\ell})],[\delta\tilde{\Xi}-\mathbf{d}I(\tilde{\ell})]\bigg)\right>,
\end{align}
where $R$ is the infinitesimal generator of gyrophase rotations times the local gyrofrequency, $I$ is the inverse of the Lie derivative $L_R$, angle brackets denote gyroangle averaging, and $\tilde{Q}=Q-\left<Q\right>$. $\mathcal{H}^{\text{gc}}$ denotes the guiding center Hamiltonian truncated at some desired order in $\rho/L$. The function $\ell$ and the $1$-form $\delta\Xi$ are defined in terms of any choice of the guiding center Lie generators as follows. Decompose the guiding center transformation $\tau_\text{gc}:TQ\rightarrow TQ$ as $\tau_{\text{gc}}=\tau_2\circ\tau_1$, where
\begin{align}
\tau_1&=\exp(G_1)\\
\tau_2&=\dots\circ\exp(G_3)\circ\exp(G_2)\equiv\exp(\bar{G}_2),
\end{align}
and the $G_k$ are the guiding center Lie generators. The $1$-form
\begin{align}
\delta\Xi=-\frac{e}{c}(\tau_{2*}\text{i}_{G_1}U_1 B+\text{i}_{\bar{G}_2}U_2 B),
\end{align}
where the lowered $*$ denotes pushforward, $U_1=\int_0^1\exp(\lambda G_1)_*\,d\lambda$, and $U_2=\int_0^1\exp(\lambda \bar{G}_2)_*\,d\lambda$, represents the gauge invariant component of the perturbation to the guiding center Lagrange $1$-form produced by the fluctuating electromagnetic fields. The function
\begin{align}
\delta\mathcal{H}=e(\tau_{2*}\text{i}_{G_1}U_1E+\text{i}_{\bar{G}_2}U_2 E)
\end{align}
represents the gauge invariant component of the perturbation to the guiding center Hamiltonian caused by the same fields. The function 
\begin{align}
\ell=\delta\Xi(V_o^{\text{gy}})-\delta H,
\end{align}
where $V_o^{\text{gy}}$ is the gyrocenter phase space velocity neglecting terms in the gyrocenter kinetic energy of order $\epsilon_\delta$ or higher, represents the gauge invariant perturbation to the guiding center Lagrangian. 

This formulation reduces to the Hamiltonian formulation of the Vlasov-Maxwell equations\,\cite{Morrison_1980,Marsden_1982} under the substitutions
\begin{align}
K&\rightarrow\frac{1}{2}mv^2\\
\Xi^{\text{gc}}&\rightarrow m v\cdot dx.
\end{align}
It is also interesting to compare $[\cdot,\cdot]_{\text{GVM}}$ to the bracket given by Morrison in Ref.\,\cite{Morrison_lifting_2013}. The only significant difference comes from the manner in which the inductive electric field is built into the kinetic equation. Finally, we note that a useful future direction of research would be identifying a Poisson bracket for electromagnetic gyrokinetics in the Darwin approximation \footnote{The Darwin approximation is typically invoked in electromagnetic gyrokinetics in order to theoretically suppress light waves. On the other hand, due to the structural similarities between the GVM equations and the standard Vlasov-Maxwell equations, it may be preferable from a numerical point of view to avoid the Darwin approximation altogether. The GVM equations have a more local field solve that is much more parallelizable than the Poisson field solves introduced by the Darwin approximation. Moreover, the polarization effects that accompany the gyrocenter transformation significantly decrease the speed of light, which leads to a more favorable CFL condition than in standard Vlasov-Maxwell simulations.}. The gyrokinetic Vlasov-Darwin equations are somtimes also referred to as the gyrokinetic Vlasov-Poisson-Amp\`ere equations \cite{Sugama_2000}. A Hamiltonian formulation of the non-gyrokinetic Vlasov-Darwin equations has already been given in Ref.\,\cite{Krauss_2007}


%
%

%

\begin{acknowledgments}
This work was supported by DOE contracts DE-AC02-09CH11466 (JWB and HQ), DE-SC0006721 (AJB), and  DE-FG05-80ET-5308 (PJM).
\end{acknowledgments}






\providecommand{\noopsort}[1]{}\providecommand{\singleletter}[1]{#1}%
%


\end{document}